\documentclass[useAMS,usenatbib]{mn2e}  
\usepackage{graphicx}
\title{Modelling the multi-wavelength emission of flat spectrum radio quasar 3C 279}
\author[Zheng \& Yang]{Y.G. Zheng$^{1}$\thanks{Corresponding author. E-mail: ynzyg@ynu.edu.cn}, C.Y. Yang$^{2,3,4}$\\
       $^{1}$Department of Physics, Yunnan Normal University, Kunming, 650092, China\\
       $^{2}$Yunnan Observatories, Chinese Academy of Sciences, Kunming 650011, China\\
       $^{3}$Key Laboratory for the Structure and Evolution of Celestial Objects, Chinese Academy of Sciences, Kunming 650011, China\\
       $^{4}$Key Laboratory of Particle Astrophysics of Yunnan Province, Kunming 650091, China}

\begin{document}
\date{Received date  / Accepted date}

\pagerange{\pageref{firstpage}--\pageref{lastpage}} \pubyear{2015}

\maketitle

\label{firstpage}

\begin{abstract}
We employ a {\bf length-dependent} conical jet model for the jet structure and emission properties of flat spectrum radio quasar 3C 279 in the steady state. In the model, ultra-relativistic leptons are injected at the base of jet and propagate along the jet structure, non-thermal photons are produced by both synchrotron emission and inverse Comtpon scattering off synchrotron photons and external soft photons at each segment of jet. We {\bf derive} the total energy spectra contribution through integrating every segment. We apply the model to the quasi-simultaneous multi-wavelength observed data of two quiescent epoches. Using the observed radio data of source, we determine the length of jet $L\sim 100$ pc, and magnetic field $B_{0}\sim 0.1-1$ G at the base of jet. Assuming a steady geometry of the jet structure and suitable physical parameters, we reproduce the multi-wavelength spectra during two quiescent observed epoches, respectively. Our results show that the initial $\gamma$-ray emission site takes place at $\sim 0.5$ pc from black hole.
\end{abstract}


\section{Introduction}
\label{sec:intro}
{\bf Blazars, a special class of active galactic nucleus (AGN), with a non-thermal continuum emission that arises from the jet emission taking place in an radio-loud AGN whose jet axis is closely aligned with the observer's line of sight(e.g. Urry \& padovani 1995; Ulrich et al. 1997), are characterized by a rapid and large amplitude variability. Multi-wavelength observations show that their broad spectral energy distributions (SED) ranging from the radio to the $\gamma$-rays bands are dominated by two components, appearing as humps. It is widely admitted that the low-energy hump that extends from radio to ultraviolet/X-ray is produced by synchrotron radiation from relativistic electrons in the jet, though the high energy hump that extends from X-ray to high energy $\gamma$-ray regime remains an open issue. In the lepton model scenarios, the high energy hump is probably attributed from invers-Compton scattering (ICs) off seed photons}. The seed photons for ICs can originate from synchrotron photons (Maraschi et al. 1992; Bloom \& Marscher 1996), accretion disk photons (Dermer et al. 1992; Dermer \& Schlickeiser 1993) and accretion disk photons re-scattered by the broad-line region clouds/interclouds medium (Sikora et al. 1994; Blandford \& Levinson 1995), or infrared radiation from a torus (Sikora et al. 1994).

{\bf Evidence shows that the energy emission of blazar originates from a region in the jet where is close to the  very long baseline interferometry (VLBI) core (e.g. Kovalev et al. 2009; Pushkarev et al. 2010; Leon-Ravares et al. 2011; Schinzel et al. 2012; Wehrle et al. 2012; Kutkin et al. 2014). In this scenarios, we could determine the properties of jet through observing the electromagnetic spectrum. The VLBI images that are observed both at large angles to the jet axis and with a spatial resolution at an order of a parsec (Pearson et al. 1996; Zensus 1997; Zensus et al. 2006; Lobanov 2010; Kutkin et al. 2014), show that there is a continuous axisymmetric plasma jet, which is a approximately conical in geometry ( Krichbaum et al. 2006; Kovalev et al. 2007) with a small degree of curvature close to the base (Asada \& Nakamura 2012) and a blunt base with a wide opening angle (Hada et al. 2011).  Measuring results of the frequency dependent core-shift are in favour of a conical jet where the magnetic flux conservation is taken into account (Sokolovsky et al. 2011). Observation indicates that the most of blazars exhibit a flat or invert radio spectra, and that their jets are characterized by a compact, bright, unresolved or bare core (e.g. Kaiser 2006; Potter \& Cotter 2012). Currently, the nature of a radio core could be interpreted as a surface in a continuous flow where the optical depth of synchrotron radiation $\tau \sim 1-$ 'photosphere' (Readhead et al. 1979; K$\rm \ddot{o}$nigl 1981; Zensus et al. 1995), However, the radiation mechanism of a flat or invert radio spectra is still an open issue.}

In order to reproduce the flat radio spectrum, a possible model was proposed by Blandford \& K$\rm \ddot{o}$nigl (1979). In the model, they {\bf adopted} a uniform conical jet structure with the conversation of magnetic flux along the jet to fit the flat radio spectrum (Blandford \& K$\rm \ddot{o}$nigl 1979). {\bf Based on the bias of observational data}, Further development of the models, {\bf where the replenishment} of adiabatic and radiative losses are deemed to be problematic, {\bf indicated} that these {\bf models} could reproduce the observed synchrotron emission spectrum(e.g. Marscher 1980; K$\rm \ddot{o}$nigl 1981; Reynolds 1982; Mufson et al 1984; Ghisellini et al. 1985). More recently, the time-dependent models were proposed to reproduce the multi-wavelength spectrum of blazar during flaring state (e.g. Kirk et al. 1998; B$\rm \ddot{o}$ttcher \& Chiang 2002; Li \& Kusunose 2000; Tsang \& Kirk 2007; Zheng \& Zhang 2011; Zheng et al. 2013; Zheng et al. 2014). In these models, The authors argue that multi-wavelength emission comes from a small spherical region of jet plasma. Assuming {\bf an} injected electron spectrum with a power-law (or broken power-law) and high energy cutoff, they {\bf calculated} the time evolution of the spectrum with electron energy losses and escaping {\bf from} blob in the Fokker-Planck equation frame. As an open issue, Above models remain an arbitrary energy injection mechanism and leave the flat radio spectrum in the future. Although an analytic model with the conical jet geometry could result in a flat radio spectrum (Kaiser 2006), unfortunately, it did not reproduce the inverse-Compton component spectrum.  Using a simply uniform conical jet, Potter \& Cotter (2012) attempted to fit multi-wavelength simultaneous spectrum of blazar, and then, including a variable geometry with an accelerating parabolic base and slowly decelerating conical jet was developed by Potter \& Cotter (2013).

A standard model that consists of magnetized plasma {\bf were} ejected with relativistic speeds in a collimated outflow along the polar axes of a rotating black hole has been developed to explain the observational features of blazar (B$\rm \ddot{o}$ttcher et al. 2012). However, even the simple lepton model {\bf suffers} from both characterizing the non-thermal electron spectrum and location of the $\gamma$-ray emission site. Motivated by above {\bf issues}, in this paper, we employ a length-dependent conical jet model to investigate the jet structure and emission properties of 3C 279 in the steady state. By reason that the SED of the flat spectrum radio quasar (FSRQ) shows the the characteristics of weak synchrotron emission, but {\bf has} stronger ICs. In the model, we consider that seed photons for ICs originate from both synchrotron photons and external radiation field. We use numerical integration of exact expression to calculate the ICs. In our calculation, we assume that the energy density of external radiation field depends on the distance between position of emission regions and core. The present work differs from the earlier studies {\bf on which adopted} a homogeneous sphere emission blob. {\bf This work focuses on the electron population that propagates along the jet structure}, and the emission region continuously changes in the fluid frame. As a simple case, we do not consider (1)the energetic transformation between magnetic field and particles in the assumption of the jet with a constant bulk Lorentz factor, (2) the additional electron population injection during in the electron evolution process, and (3) the extragalactic background light (EBL) absorption. Throughout the paper, we assume the Hubble constant $H_{0}=75$ km s$^{-1}$ Mpc$^{-1}$, the matter energy density $\Omega_{\rm M}=0.27$, the radiation energy density$\Omega_{\rm r}=0$, and the dimensionless cosmological constant $\Omega_{\Lambda}=0.73$.

\section{Model Description}
{\bf The photon spectra in the context are reproduced by the model within the lepton model frame through both synchrotron self-Compton (SSC) emission and external Compton (EC) emission. In the model, We basically follow the approach of Potter \& Cotter (2012) to reproduce SSC emission(for more details, see Potter \& Cotter 2012), and we introduce an external radiation field that includes both broad lines region (BLR) and infrared dust emission (IR) to the EC emission. With the aim of readability, A brief description of the model is provided as follows.

\subsection{The SSC Photon Spectra}
The model that we adopt assumes relativistic plasma propagates with an associated bulk Lorentz factor $\Gamma$ in a stationary funnel and the geometry is a truncated cone of length $\Gamma L$. We define the dynamic variable $x$ as the length along the jet axis in the fluid frame, where $x=0$ is the base of the jet and $x=L$ is the end of the jet. We can now parameterize the geometry of the jet as follows:
\begin{equation}
R(x) = R_{0} + x \tan \theta_{\mathrm{opening}} .
\end{equation}
where $R(x)$ is the radius of the jet at the length $x$, $R_{0}$ is radius at the base of jet, and $\theta_{\mathrm{opening}}$ is a half opening angle of the cone.

Using the length-dependent SSC solution for above geometry, we can reproduce multi-wavelength photon spectra. In order to do so, we first determine both the radius $R_{0}$ and a initial injected electron distribution $N_e(E_e,0)$ with the cut off energy $E_{\rm e, cut}$ at the base of jet.
\begin{equation}
R_0=\sqrt{\frac{2E_{j}A_{\rm{equi}}\mu_0}{\Gamma^{2}(\pi B_0^2)(1+A_{\rm {equi}})}},
\end{equation}
where $E_j$ is the energy contained in a section of plasma of width 1 m in the $x$-direction in the lab frame which is equal to $W_j/c$ with total jet power $W_j$ in the lab frame, $A_{equi}$ is the equipartition fraction between magnetic field energy $U_B$ and electron energy $U_e$, $B_{0}$ is magnetic field at the base of jet, and $\mu_{0}$ is magnetic permeability.
\begin{equation}
N_e(E_e,0)\simeq A_{0}E_e^\alpha \rm{e}^{-E_e/E_{\rm e, cut}}
\end{equation}
where, $A_{0}\simeq(2-\alpha)E_{j}/[\Gamma^{2}(1+A_{equi})(E_{\rm e, cut}^{2-\alpha}-E_{e, min}^{2-\alpha})]$ (Potter \& Cotter 2012) with a minimum electrons of energy $E_{e, min}$, and $\alpha$ is electron spectra index ranging between 1 and 3 from the theory of shock acceleration (Bell 1978; Bell et al. 2011; Summerlin \& Baring 2011).

We now consider the evolution properties of the magnetic field $B(x)$ and electron spectra $N_e(E_e,x)$. Assuming the magnetic energy conservation and particle density conservation in each segment, we calculate the length evolution of the the magnetic field $B(x)=B_{0}R_{0}/R(x)$ and electron spectra $N_e(E_e,x+dx)=N_e(E_e,x)-P_{tot}(x,dx,E_e)/cE_{e}$ to the end of the jet, where $P_{tot}$ is the total power emitted by electrons of energy $E_e$ by a section of jet of width $dx$ in fluid frame due to synchrotron emission and ICs. Therefore, the total SSC photon spectra in the fluid frame is the summation of the synchrotron emission and ICs from all the segment in the jet.}

\subsection{External Radiation Field}

As described above, {\bf the contribution of the external radiation field that includes both broad lines region (BLR) and infrared dust emission (IR) are considered}. The external radiation field energy densities in the jet comoving frame as function of the distance $r$ along the jet are approximated by formulae (Hayashida et al. 2012):
\begin{equation}
u_{\rm {BLR}}(r)=\frac{\eta_{\rm {BLR}}\Gamma^2 L_D}{3\pi r_{\rm{BLR}}^2c[1+(r/r_{\rm{BLR}})^{\beta_{\rm{BLR}}}]},
\label{e:8}
\end{equation}
and
\begin{equation}
u_{\rm{IR}}(r)=\frac{\eta_{\rm{IR}}\Gamma^2 L_D}{3\pi r_{\rm{IR}}^2c[1+(r/r_{\rm{IR}})^{\beta_{\rm{IR}}}]},
\label{e:9}
\end{equation}
where $\eta_{\rm {BEL}}$ and $\eta_{\rm {IR}}$ are the fractions of the disk luminosity $L_D$ reprocessed into broad lines region and into hot dust radiation, respectively, $r_{\rm{BLR}}\simeq0.1(L_{D}/10^{46}\rm erg~s^{-1})^{1/2}$ pc and $r_{\rm{IR}}=2.5(L_{D}/10^{46}\rm erg~s^{-1})^{1/2}$ pc are the characteristic distance where the above reprocessing takes place. The external radiation fields in the jet comoving frame by Maxwellian distribution peaked at photon energies $E_{\rm BLR}\sim 10 ~\rm eV\times \Gamma$ and $E_{\rm IR}\sim 0.3 ~\rm eV\times \Gamma$. In our calculation, we adopt the radiation density profile $\beta_{\rm BLR}=3$ (Sikora et al. 2009) and $\beta_{\rm IR}=4$ (Hayashida et al. 2012), respectively.

Because {\bf the initial $\gamma$-ray emission site that is defined by $x=0$ in the model} does not show the real distance between the core and base of jet, we introduce another parameter $L_{B}$ to describe it in the fluid frame. Then, the distance $r$ along the jet in equation (\ref{e:8}) and (\ref{e:9}) could be determined by $r=L_{B}+x$. Although the parameter $r$ shows the distance between radiation region and core in the jet comoving frame (e.g. Hayshida et al. 2012; Dermer et al. 2014), the emission region moves along {\bf $x$-axis of jet} continuously. In our calculation, we derived the total ICs contribution through integrating every segment from $x=0$ to $x=L$.

\section{Apply to 3C 279}

The distant FSRQ 3C 279, at a redshift of $z=0.536$, is a strong $\gamma$-ray source that was discovered by EGRET in 1991 (Hartman et al. 1992). Multi-wavelength snapshot observations for several {\bf epoches} including $\gamma$-rays show that the SED could be explained by the leptonic model (Hartman et al. 2001). In order to comprehend intensively the energy spectrum character of the source, a multi-wavelength quasi-simultaneous observation campaign has been organized by Hayashida et al. (2012). Now that the aim of the present work is to study in more detail the jet structure and emission properties of 3C 279 in the steady state. {\bf In this paper}, we only use the data of two quiescent epoches, that is, during Aug-Sep in 2008 (A state) and during Feb-May in 2010 (B state). The SEDs comprise X-ray data from {\it Suzaku}, {\it Swift} XRT, {\it XMM Newton}, and {\it RXTE}; optical/UV data from Kananta, GROND, and {\it Swift} UVOT (170-650 nm); IR data from {\it Spitzer}, and radio data from CARMA and OVRO; $\gamma$-ray data from {\it Fermi} LAT.

\subsection{Constraint on the Length and Magnetic of Jet}
\begin{figure}
	\centering
		\includegraphics[angle=0,width=9 cm]{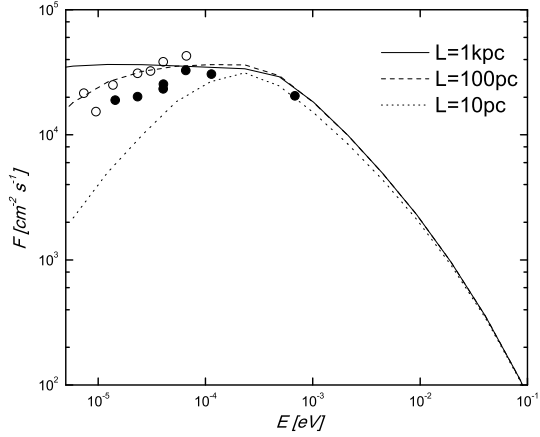}
		
	\caption{The differential synchrotron flux with differen the length of jet ($L$). The blank and solid circle are the radio data of A and B state respectively. The other input parameters were fixed to refer to A state of Table \ref{Table:1}. }
	\label{fig:1}
\end{figure}

The observation of a flat/reversed radio spectrum is a characteristic of most blazars (e.g. Abdo et al. 2010). {\bf Given that} the model parameter $L$ and $B_{0}$ dominate on the energy range of flat/reversed radio spectrum (Potter \& Cotter 2012), we attempt to determine the length and magnetic of jet through the observed radio data. In order to do that, in Figure $\ref{fig:1}$, we first calculate the radio differential flux spectra with different length of jet $L$. For comparison, the observed radio data of 3C 279 at two quiescent epoches (Hayashida et al. 2012) are also shown in Figure $\ref{fig:1}$ (blank and solid circle). It can be seen that (1) a flat radio spectrum is largely insensitive to the length of jet, and that it mainly affects the extend of the flat radio spectrum to low frequencies; (2) the observation of a flat/reversed radio spectrum is reproduced quite well with the length of the jet $L\sim 100$ pc for the source; (3) the energy of departure from the flat/reversed spectrum locates on around $\sim 10^{-4}$ eV. Although, {\bf the investigation focuses on} the observed data of two quiescent epoches, we actually should derive the same results even during in flare epoches.

\begin{figure}
	\centering
		\includegraphics[angle=0,width=9 cm]{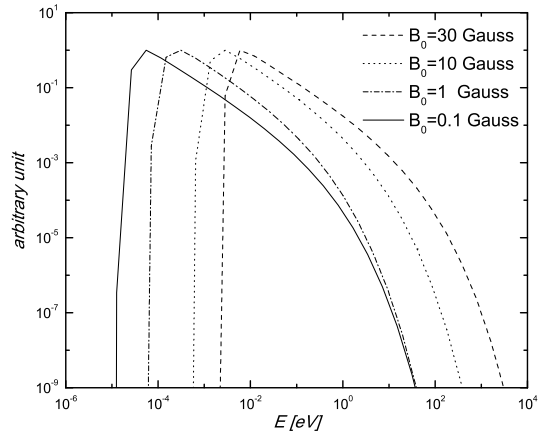}
		
	\caption{The synchrotron emission flux at different magnetic field ($B_0$) at the base of jet ($x=0$), the flux has been normalized the maximum value for each value of the magnetic field. The other input parameters were fixed to refer to A state of Table \ref{Table:1}.}
	\label{fig:2}
\end{figure}

We now search for the expected energy ranges of high frequency departure from the flat spectrum. In order to do it, we show the predicted synchrotron emission flux with different magnetic field $B_0$ at the base of jet in Figure $\ref{fig:2}$. In the figure, the emission flux has been normalized using the maximum flux with the adopted magnetic field $B_0$. The result shows a tendency of that the reversed energy increase when the magnetic field strengthen at the base of jet. Because that there is a relation between the magnetic field of the jet at the length $x$ and the length of jet (see Eq. (3)), we argue that the reversed energy at the base of jet should be the energy of departure from the flat/reversed spectrum. If {\bf above} issue that the energy of departure from the flat/reversed spectrum locates on around $\sim 10^{-4}$ eV is true, we could give a constraint on the magnetic field $B_0\sim 0.1-1$ G at the base of jet.

\subsection{Modelling the Multi-wavelength Spectra}
Using the synchrotron and ICs solution for the conical jet structure, we can calculate multi-wavelength spectra in two quiescent epoches, respectively. In order to do so, we search for the electron and photon spectra along {\bf $x$-axis of jet}. Since we consider that seed photons for ICs originate from both synchrotron photons and external radiation field, we should calculate the emission spectrum from accretion disk. In our calculation, we adopt a simplified Shakura-Sunyaev disk spectrum (Dermer et al. 2014) as follows:
\begin{equation}
\epsilon L_D (\epsilon)=1.12 L_D(\frac{\epsilon}{\epsilon_{max}})^{4/3}\exp(-\frac{\epsilon}{\epsilon_{max}})
\end{equation}
where $\epsilon$ is the photon energy that is emitted by accretion disk. While the value of $\epsilon_{max}$ depends on the spin of the black hole and relative Eddington luminosity, we set is as a typical characteristic temperature of UV bump in Seyfert galaxies with $\epsilon_{max}\sim 10$ eV.

Assuming a steady geometry of the jet structure with $\theta'_{\mathrm{opening}}=3.0^{o}$, $L=100$ pc, and a immovable position of the observer with $\theta_{\mathrm{observe}}=2.0^{o}$ during in two different observed epoches, we calculate the electron and photon spectrum in every {\bf segment} from $x=0$ to $x=L$. {\bf We argue that the photons with energy from X-ray to $\gamma$-ray are produced by ICs off external radiation field. Therefore, the disk luminosity $L_D$ could be determined by fitting the observed data from X-ray to $\gamma$-ray}. For two observation epoches of 3C 279, we adopt $L_D=6\times 10^{45} \rm erg~s^{-1}$ and $L_D=4\times 10^{45} \rm erg~s^{-1}$, respectively. {\bf These values are in consistent with an accretion-disk spectra luminosity $L_D\approx 2\times 10^{45} \rm erg s^{-1}$ that are found from excess optical/UV radiation} (Pian et al. 1999). {\bf the parameters are listed in table $\ref{table1}$}. We note that the jet opening angle $\theta^\prime_{\rm opening}$ in table $\ref{table1}$ in the lab frame is related to the fluid frame opening angle $\theta_{\rm opening}$ via $\Gamma \tan \theta^\prime_{\rm opening}=\tan \theta_{\rm opening}$.

\begin{table*}
\begin{minipage}[t][]{\textwidth}
\caption{The physical parameters of the uniform conical jet model spectra}
\label{table1}
\begin{tabular}{lll}
\hline\hline
 Parameter & A state & B state \\
\hline\\
$W_{j}$ & $1.7 \times 10^{40}~\rm{W}$ & $2.6 \times 10^{40}~\rm{W}$\\
L & $100 ~\rm{pc}$ & $100 ~\rm{pc}$ \\
$B_{0}$ & $0.15~\rm{G}$ &  $0.15~\rm{G}$\\
$L_{B}$ & $0.5$ pc & $0.5$ pc \\
$A_{\rm{equi}}$ & 0.015 & 0.015 \\
$E_{min}$ & 5.11 \rm{MeV} & 5.11 \rm{MeV}\\
$E_{cut}$ & 1.0 \rm{GeV} & 0.3 \rm{GeV}\\
$L_D$ & $6.0 \times 10^{45}$ $\rm{erg~s^{-1}}$ & $4.0 \times 10^{45}$ $\rm{erg~s^{-1}}$ \\
$\eta_{\rm{BLR}}$ & 0.2 & 0.1 \\
$\eta_{\rm{IR}}$ & 0.4 & 0.3 \\
$\alpha$ & 2.6 & 2.0 \\
$\theta'_{\rm{opening}}$ & $3.0^{o}$ & $3.0^{o}$ \\
$\theta_{\rm{observe}}$ & $2.0^{o}$ & $2.0^{o}$ \\
$\Gamma$ & 8 & 15\\
\hline\\
\end{tabular}
\end{minipage}
\end{table*}

\begin{figure}
	\centering
		\includegraphics[angle=0,width=9cm]{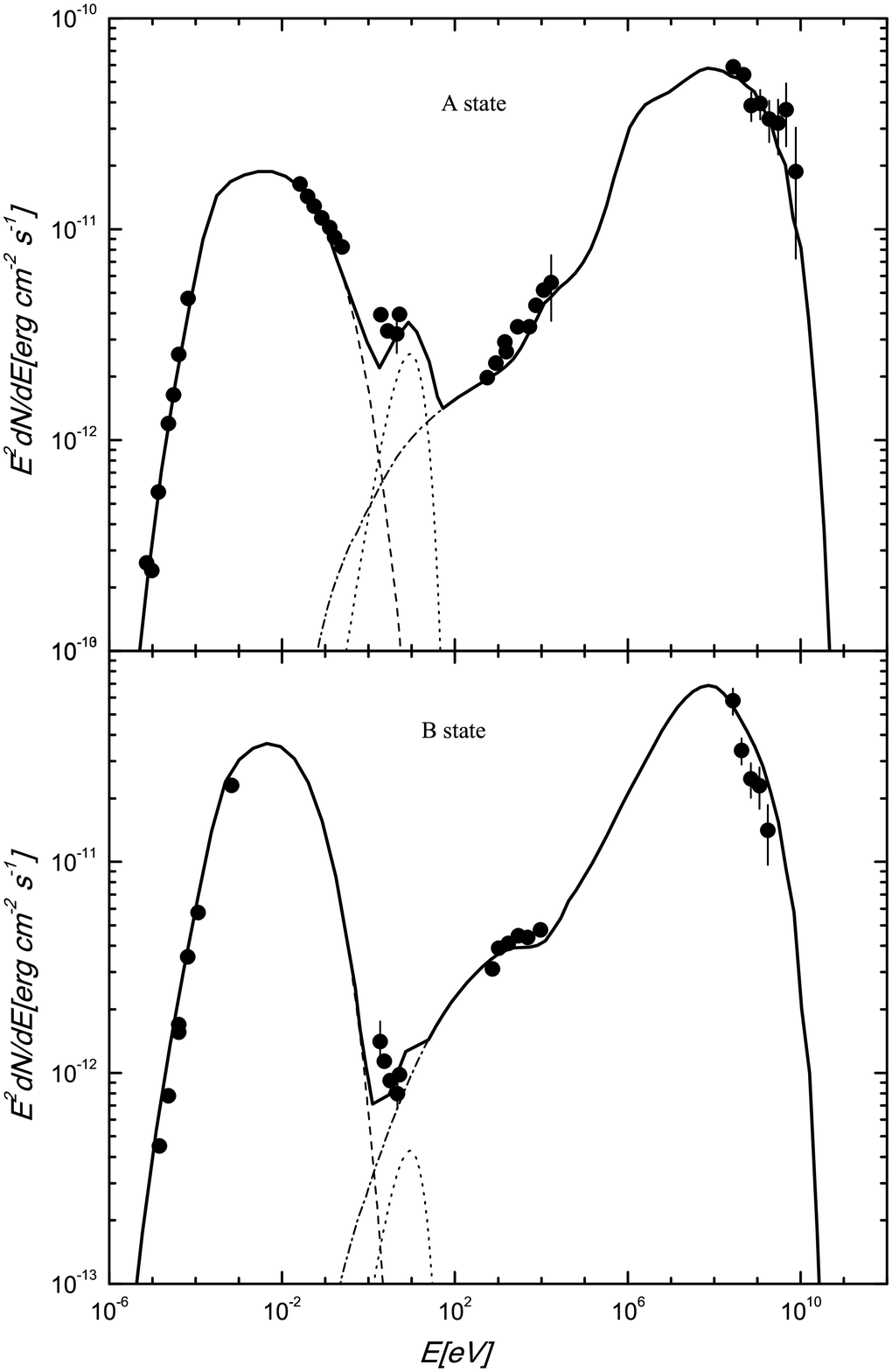}
	\caption{Comparisons of predicted multi-wavelength spectra with observed data of 3C 279 during in Aug-Sep in 2008 (top panel) and during in Feb-May in 2010 (bottom panel). The thin solid line represents the synchrotron emission, the dashed line represents ICs on the seed photons of synchrotron, BLR and IR, the dotted line represents the radiation from accretion-disk, and the thick solid line represents total spectrum by summation all of emission components, respectively. Observed data come from Hayashida et al. (2012).}
	\label{fig:3}
\end{figure}

We assume that observed spectrum is a summation of each segment. Therefore, we can calculate the observed spectrum in different observed epoches using the photon spectrum in every segments from $x=0$ to $x=L$. In Figure $\ref{fig:3}$, we show predicted multi-wavelength spectrum (the solid curve). For comparison, the observed data of 3C 279 at two quiescent epoches (Hayashida et al. 2012) are also shown, where black solid circles with error bars represent the observed values. It can be seen that the observed data in the pre-burst state can be reproduced in the model.

\subsection{Location of $\gamma$-ray Emission Site}

Traditionally, a simple emission model {\bf considers} that high energy photon comes from a blazar-like blob, where the distance to the core is a fixed value. We could not locate the $\gamma$-ray emission site according to a adopted size of blob. In the present model, as a free parameter, the distance between the core and the base of jet $L_{B}$ determine the energy density of BLR and IR at the base of jet, and the intensity of external Compton component. Because that we derived the total ICs contribution through integrating every segment from $x=0$ to $x=L$, and we change the distance between $\gamma$-ray emission site to core continuously from $L_{B}$ to $L_{B}+L$ in fluid frame. Therefore, we could use the parameter $L_{B}$ to define minimum radii for $\gamma$-ray production, that is, the parameter $L_{B}$ could trace the initial location of $\gamma$-ray emission site from black hole. In order to reproduce the observed $\gamma$-ray photon spectra, in our calculation, we adopt $L_{B}=0.5$ pc. This {\bf finds} that the initial $\gamma$-ray emission site takes place at $\sim 0.5$ pc from black.

\section{Discussion and Conclusion}
It is considered that the multi-wavelength emission of a blazar encodes important information about particle acceleration and radiation, jet structure and environment. In order to determine the jet structure and the location of $\gamma$-ray emission site in blazar, there is great interest in the inversion of the jet physics and environment from the SED of blazar (e.g. Bloom et al. 2013; Dermer et al. 2014). In this paper, we employ a {\bf length-dependent} conical jet model for the jet structure and emission properties of FSRQ 3C 279 in the steady state. In the model, ultra-relativistic leptons are injected at the base of jet and propagate along the jet structure, non-thermal photons are produced by both synchrotron emission and inverse Comtpon scattering off synchrotron photons and external soft photons at each segment of jet. We {\bf derive} the total energy spectra contribution through integrating every segment. We apply the model to the quasi-simultaneous multi-wavelength observed data of two quiescent epoches. Using the observed radio data of source, we determine the length of jet $L\sim 100$ pc, and magnetic field $B_{0}\sim 0.1-1$ G at the base of jet. Assuming a steady geometry of the jet structure and suitable physical parameters, we reproduce the multi-wavelength spectra during two quiescent observed epoch, respectively. Our results show that the initial $\gamma$-ray emission site takes place at $\sim 0.5$ pc from black hole.

The model presented in this work {\bf considers} a more complex geometry of the jet structure than a homogeneous sphere emission blob. In the model, particle spectra are calculated self-consistently according to the given geometry and assumed physical conditions. Therefore, instead of the spectral parameters, we have to describe jet structure and the processes that control the evolution of the spectrum. Furthermore, we consider the contribution of the external radiation field, the external radiation field energy densities in the jet comoving frame should be introduced. In principle, the model requires thirteen free parameters($L$, $\theta_{\mathrm{opening}}$, $\Gamma$, $A_{\rm equi}$, $B_{0}$, $W_{j}$, $\theta_{\mathrm{observe}}$, $E_{min}$, $E_{cut}$, $\alpha$, $L_{B}$, $L_{D}$, and $\eta_{BLR}$ or $\eta_{IR}$). In this particular test, due to the model parameter $L$ and $B_{0}$ {\bf dominating} on the energy range of flat/reversed radio spectrum (Potter \& Cotter 2012), we could determine the length $L$ and magnetic of jet $B_{0}$ through the observed radio data. {\bf Therefore, the number of free parameter is reduced to eleven}. Generally, {\bf the contribution of the external radiation field could not be taken into account for modeling the SED of BL Lac object (e.g. Katarzynski et al. 2006), we reduces the free parameters to eight. This is more one free parameter than the number of free parameter that is required by a simple stationary synchrotron self-Compton scenario}.

Dependence on the quasi-simultaneous multi-wavelength observed data of two quiescent epoches, the model also {\bf predicts} that the initial $\gamma$-ray emission site takes place at $\sim 0.5$ pc from nucleus, though it changes with the accretion disk luminosity $L_{D}$. This location of the $\gamma$-ray emission site is consistent with relativistic shell collision radii that is parameterized within the colliding shell paradigm for blazar(e.g. Spada et al. 2001; B$\rm \ddot{o}$ttcher \& Dermer 2010; Mimica \& Aloy 2012). As a result, it is indicated that the {\bf possible} location of the emission region is near the BLR ($r_{BLR}\sim 0.1$ pc) and deepest within IR tours ($r_{IR}\sim 1.6$ pc). The inferred initial $\gamma$-ray emission site at same scale in two different observed {\bf epoches is} easier to explain for a steady geometry of the jet structure with different accretion disk radiation during quiescent epoches. Now that scattered accretion disk emission (e.g. Hartman et al. 2001; Dermer \& Schlickeiser 2002) {\bf provides} a similar emission component with IR emission, we argue that it is important to determine if the $\gamma$-ray emission region is made deep within the BLR or farther out, where the density of BLR is more dilute. We leave this open issue to the studies on $\gamma\gamma$ attenuation at very high energy $\gamma$-ray regime (e.g. Liu \& Bai 2006; Sitarek \& Bednarek 2008).

As a critical physical parameter, the magnetic distribution $B(x)$ could determine the shape of radio spectrum (Potter \& Cotter 2012) that is reproduce by the model. It is well {\bf known} that observation of a flat/reversed radio spectrum is a characteristic of most blazars. In order to reproduce a flat radio spectrum, in {\bf the} present work, we assume that the magnetic energy is conserved along the jet. This assumption results to a pure toroidal magnetic field distribution in the jet, while a polodial magnetic distribution is adopted in an inhomogeneous jet model for the rapid variability of TeV blazars (Boutelier et al. 2008). We note that a dominant toroidal magnetic field in jet are predicted by theory (Lyutikov et al. 2005) and magnetohydrodynamic (MHD) simulations (McKinney 2006; Komissarov et al. 2007), and are also supported by a parsec-scale polarization observation (O'Sullivan \& Gabuzda 2009; Kharb et al. 2009).

A potential drawback of the model is that we do not specify the acceleration mechanism. As a simple case, we assume that ultra-relativistic leptons are injected at the base of jet and propagate along the jet structure. Since we consider a variational magnetic field in each segment, the distribution of moving magnetic inhomogeneities could result to a Fermi-type acceleration (Fermi 1949; 1954). In this scenario, a more complex particle spectra than the model presented in this work should be expected. Although, It is a simple model in the context, using the quasi-simultaneous multi-wavelength observed data of the source, we could determine both the length of jet and magnetic field at the base of jet, and infer the initial $\gamma$-ray emission site in FSRQ. We {\bf recommend} that the model should be used to the inversion of the jet physics and environment from the SED of blazar.

\section*{Acknowledgments}
We thank the anonymous referee for valuable comments and suggestions.
This work is partially supported by the National Natural Science Foundation of China under grants 11463007, U1231203, the Strategic Priority Research Program ``the Emergence of Cosmological Structures" of the Chinese Academy of Sciences (Grant XDB09000000), Science and Technology in support of Yunnan Province Talent under grants 2012HB014, and the Natural Science Foundation of Yunnan Province under grant 2013FD014. This work is also supported by the Key Laboratory of Particle Astrophysics of Yunnan Province (grant 2015DG035).


\end{document}